\documentclass[3p, times]{elsarticle}
\usepackage{graphicx}
\usepackage{color}
\usepackage{amssymb}
\usepackage{amsmath}

\usepackage [english]{babel}
\usepackage [autostyle, english = american]{csquotes}
\MakeOuterQuote{"}

\begin{document}

\title{Entropy and the second law for  driven, or quenched, thermally isolated systems}
\author{ Udo Seifert }
\ead{useifert@theo2.physik.uni-stuttgart.de}

\address{ {II.} Institut f\"ur Theoretische Physik, Universit\"at Stuttgart,
  70550 Stuttgart, Germany\\~\\ }

\begin{abstract}

	The entropy of a thermally isolated system should not decrease after
	a quench or external driving.
	For a classical system following Hamiltonian dynamics,
	we show how this statement emerges for a large system in the
	sense that the extensive part of the entropy change does not become negative.
	However, for any finite system and small driving, the mean entropy change can well be 
	negative.
	We derive these results using as micro-canonical entropy a variant
	recently introduced by Swendsen and co-workers called "canonical".
	This canonical entropy is 
	the one of a canonical ensemble with the corresponding mean energy.
	As we show by refining the micro-canonical Crooks relation, the same
	results hold true for the two more conventional choices of micro-canonical
	entropy given either by the area of a constant energy shell,
	the Boltzmann entropy,
	or the volume underneath it, the Gibbs volume entropy. These results are 
	exemplified with quenched $N$-dimensional harmonic oscillators.


\end{abstract}


\maketitle

\def\l{\lambda}

\def\beq{\begin{equation}}

\def\lsim
{\protect \raisebox{-0.75ex}[-1.5ex]{$\;\stackrel{<}{\sim}\;$}}
\def\gsim
{\protect \raisebox{-0.75ex}[-1.5ex]{$\;\stackrel{>}{\sim}\;$}}
\def\lsimeq
{\protect \raisebox{-0.75ex}[-1.5ex]{$\;\stackrel{<}{\simeq}\;$}}
\def\gsimeq
{\protect \raisebox{-0.75ex}[-1.5ex]{$\;\stackrel{>}{\simeq}\;$}}

\def\beq{\begin{equation}}
\def\ee{\end{equation}}

\def\bi{\begin {itemize}}
\def\ei{\end{itemize}}

\def\pcite{\protect\cite}

\def\nn{\nonumber}


\def\lsim
{\protect \raisebox{-0.75ex}[-1.5ex]{$\;\stackrel{<}{\sim}\;$}}

\def\gsim
{\protect \raisebox{-0.75ex}[-1.5ex]{$\;\stackrel{>}{\sim}\;$}}

\def\lsimeq
{\protect \raisebox{-0.75ex}[-1.5ex]{$\;\stackrel{<}{\simeq}\;$}}

\def\gsimeq
{\protect \raisebox{-0.75ex}[-1.5ex]{$\;\stackrel{>}{\simeq}\;$}}

\def\xib{{\boldsymbol \xi}}
\def\H{{\cal H}}

\def\bS{{S_C}}
\def\bb{\beta_C}
\def\bbb{\beta_B}
\def\DS{\Delta{\cal S}}
\def\Smic{S_{B}} 
\def\Sg{S_{G}}

\def\xib{\boldsymbol \xi}
\def\lb{{\lambda}}

\def\bxi{\xib}
\def\xib{\boldsymbol \xi}

\def\W{{\cal W}}
\section{Introduction}
A main purpose of the concept of entropy is to rationalize the second law.
In one of its
incarnations it says that in a driven isolated system, entropy should not decrease.
This scenario includes the  more restricted cases of releasing
a constraint or establishing  contact between previously separated systems,
since both processes can be realized by  a sudden change of a control parameter in the total Hamiltonian, i.e., by a quench.  
Even if one accepts the crucial assumption that an isolated system finally equilibrates, 
as we will do throughout this paper, there remains the task to show that the entropy
does not decrease.
   
 The quest for a proof of such a dynamical version of the second law involving
 entropy should conceptually be distinguished
 from apparently related results for second law-like statements
 involving the mean work spent or extracted in such processes for which one must
 distinguish two types of initial conditions. For sampling from a canonical distribution,
 one can prove the Kelvin-Planck statement of the second law asserting that
 work cannot be extracted from a cyclic variation even for a finite system \cite{bass78,lena78,jarz97,alla02}.
 On the other hand,
  for micro-canonical initial conditions,  there are explicit examples of low-dimensional
  systems that demonstrate the contrary \cite{sato02a,mara10}. 
  Crucially for the present context, these, and related  \cite{kawa07}
  results about work  do not require, or invoke,
 any notion of entropy and hence do not imply a second law for entropy without 
 additional assumptions.

 Various forms for the entropy for an isolated system in equilibrium have recently been discussed,
 compared and criticised, see \cite{hilb14,swen17} for review-like presentations.
 One standard candidate for entropy under micro-canonical initial conditions is the one derived from the
size of a thin energy shell, often called Boltzmann entropy. One of its
obvious short-comings is the formal necessity to introduce an ill-defined thickness of the
energy shell to get the dimensions correct. A complementary definition often called Gibbs volume entropy that involves all states below a certain energy
avoids this fuzziness. Recently, introducing a third variant,
 Swendsen and co-workers have suggested to assign to
an isolated quantum system the standard entropy of that canonical ensemble that has as mean energy 
the originally given micro-canonical energy \cite{swen15,matt17}. We will adapt
this definition to classical systems with unbounded kinetic energy that are of interest here. 
 While for a large system with a monotonically increasing density of states,
all three variants become equivalent, they differ in the extent to which they obey, 
for finite systems or systems with a non-monotonic density of states, the various
axioms of thermodynamics.
Quoting recent representative  work emphasizing the merits of the Boltzmann entropy 
\cite{vila14,fren15,abra17}, Gibbs volume 
entropy \cite{dunk14,camp15a}, canonical entropy  \cite{swen15,matt17} and
still another one \cite {fran18} should suffice to get an entry to this debate,
which is not the main topic of this  work.

The main purpose of the present paper is rather to explore the potential of the three
variants, and, in
particular, that of canonical entropy as the least well-known for deriving a second law
for driven systems from a dynamical perspective. Assuming equilibration in an isolated
system and Hamiltonian dynamics, we first prove that the canonical entropy change evaluated at the canonical average of the final energy is non-negative for a finite system and micro-canonical initial conditions. With minor additional
assumptions about the scaling of the cumulants of work for large systems, this theorem shows how a second law for the canonical entropy change of a large system emerges
 in the sense that, for a large system, an extensive negative entropy
change can be shown not to exist. 
 However, for small driving the mean
entropy change can well become negative for any fixed system size.
Second, starting from
the micro-canonical Crooks relation \cite{cleu06a} and refining it, 
we explore the emergence of a second law for large systems also
explicitly for the Boltzmann and the Gibbs entropy.

For the present crucial micro-canonical initial conditions,
earlier work in this direction includes the observation that
for quenched one-dimensional non-linear oscillators the Gibbs volume entropy
can decrease \cite{papa11,andr14}. For large systems, 
 Sasa and Komatsu show first that for a small quench the change in volume entropy is not extensively negative. 
In a second step, using  concepts from chaotic dynamics, they conclude the same property 
for an arbitrary (most probable) process \cite{sasa00}. Tasaki proves an increase in volume entropy
in the large $N$-limit \cite{tasa00}.
For quenching or driving from canonical initial conditions, an increase in entropy can be shown even for a finite system using
the Gibbs volume entropy \cite{tasa00,camp08b}. 

It should be emphasized that the present approach and the results 
derived below are not in conflict with the well-known fact
 that under Hamiltonian dynamics a fine-grained  non-equilibrium entropy of the form $-\int d\xib p(\xib,t)\ln p(\xib,t)$
 is conserved due to Liouville's theorem. 
 Starting with such a fine-grained entropy, second law-like statements can still be obtained, either by a final weak coupling to an ideal bath
 \cite{parr09}, or by breaking dynamically induced correlations between
 system and bath \cite{espo10f}. Even for a system strongly coupled
 to a heat bath, a second law can then be proven for the sum of a suitably identified system entropy change and the entropy change of 
the bath due to the exchanged heat \cite{seif16}, see also
\cite{mill17,stra17,jarz17,aure18}.
 In the present paper, we deal with a more coarse-grained entropy function  that depends only on energy and
 the control parameters and that can be defined as an equilibrium property. 
 As is well-known, demonstrating explicitly a final equilibration 
 after a quench or driving is a quite different, much more difficult issue not addressed here.

The paper is organized as follows. In Sect. II, we state the problem.
In Sect. III, we recall the definition of canonical entropy
and derive the corresponding second law. In Sect. IV, starting with the micro-canonical Crooks
relation, we analyze the status of a dynamical second law for the
two standard variants of micro-canonical entropy. In Sect. V, 
two variants of 
 quenched harmonic oscillators are presented as specific examples. We conclude in Sect. VI.

\section{The problem}
We consider an isolated system characterized by a  $N$-particle Hamiltonian $H(\xib,\lb)$
that depends on the degrees of freedom $\xib$ and a 
 control parameters $\lb$ through which
we drive the system  for a finite time
$0\leq\tau\leq t$ from $\lb^0$ to $\lb^1$
leading to a trajectory $\xib^\tau$ in phase space. The phase point at 
the final time $t$ becomes a
function of the initial one, $\xib^1(\xib^0)\equiv \xib^t(\xib^0)$.
The work 
\beq
W(\xib^0)\equiv H(\xib^1(\xib^0),\l^1)-E^0 
\ee spent in this process is the total energy change with
$E^0 =H(\xib^0,\l^0)$.
A necessary condition for a second law to hold under this driving is that there exists
an entropy $S(E,\lb)$ such that
\beq
\langle S(E^0+W(\xib^0)),\lb^1)\rangle - S(E^0,\lb^0)\geq 0
\label{eq:putsec}
\ee for any arbitrary but fixed driving protocol $\lb^\tau$.
Throughout, averages $\langle ...\rangle$ are over micro-canonical initial conditions, i.e., over the
initial energy shell with energy $E^0$.

In the following, we explore the status of such a putative second law both for finite systems and
in the thermodynamic limit using three version of entropy.
\section{Second law for canonical entropy}
\subsection{Definition of canonical entropy}

For the isolated system with energy $E$ and a  Hamiltonian $H(\xib,\lb)$, "canonical" entropy
$\bS(E,\lb)$ is defined through considering a fictitious canonical ensemble at a
 temperature that would lead to a mean energy $E$ \cite{swen15}. Specifically,
this  substitute, or "canonical",  (inverse)  temperature  $\bb(E,\lb)$ follows from solving the implicit equation
\beq
E=-\partial_\beta\ln Z(\beta,\lb)_{|\beta=\bb(E,\lb)}
\label{eq:bb} 
\ee for $\bb(E,\lb)$. Here, the canonical partition function is given by the usual
\beq
Z(\beta,\lb)\equiv \int d\xi\exp[-\beta H(\xi,\lb)]\equiv \exp[-\beta F(\beta,\lb)] ,
\ee with the free energy $F(\beta,\lb)$. The integration is over all phase space with 
normalization factors like Planck's constant and $N$ factorials notationally suppressed.
Once $\bb(E,\lb)$ is obtained from (\ref{eq:bb}), which has a unique solution for particle-based
 system
with their monotonically increasing density of states, the canonical entropy (with Boltzmann's 
constant set to 1 throughout)
is defined as
\beq
S_C(E,\lb)\equiv \bb(E,\lb)[E-F(\bb(E,\lb),\lb)]= S(\bb(E,\lb),\lb) .
\label{eq:bS}
\ee
Here,
\beq
S(\beta,\lb) \equiv \beta^2\partial_\beta F(\beta,\lb)=\beta(U-F)
\ee
is the standard entropy of the canonical ensemble leading to a
 mean energy $U=\partial_\beta(\beta F)$. 
The relation
\beq \partial_E\bS(E,\lb)=\bb(E,\lb)
\label{eq:bes}
\ee is easily verified and
corresponds to the well-known thermodynamic one.

The change  in canonical entropy associated with
driving the system through $\lb^\tau$, after final equilibration at constant $\lb^1$, 
becomes with (\ref{eq:bS}) 
\beq
\Delta \bS(\bxi^0) = \bS(E^0+W(\bxi^0),\lb^1)-\bS(E^0,\lb^0) ,
\ee
which still depends on the initial $\xib^0$.
Our first aim is to explore under which conditions the mean
entropy change can be shown to be non-negative. 
It will be convenient to rewrite
$\Delta \bS$ as
\beq
\Delta \bS(\bxi^0) =\bS(E^0+\W +\delta W(\xib^0) ,\lb^1)-\bS(E^0,\lb^0) 
\label{eq:bbS}
\ee
with
\beq
\delta W(\xib^0)\equiv W(\bxi^0)-\W.
\ee
Here, $\W$ is defined as the mean work one would spend in such a process in a fictitious ensemble
where the initial points are drawn from a
canonical distribution at inverse  temperature $\bb(E,\lb^0)$. 
Specifically,
\beq
\W\equiv\int d\xib^0\exp\{-\bb[H(\xib^0,\lb^0)-F(\bb,\lb^0)]\}
W(\xib^0).
\ee
Expanding the first term of (\ref{eq:bbS}) in  $\delta W(\xib^0)$ 
and using (\ref{eq:bes}) leads after averaging to
\beq
\langle \Delta \bS\rangle
=\DS +\langle \delta W\rangle  \bb(E^0+\W,\lb^1)
+
\sum_{k=2}^\infty (1/k!)\langle (\delta W(\xib^0))^k\rangle \partial^{k-1}_E\bb(E,\lb^1)_{|E=E^0+\W} .
\label{eq:exp}
\ee
Here, the leading term
\beq
\DS\equiv \bS(E^0+\W ,\lb^1)-\bS(E^0,\lb^0) 
\ee 
 corresponds to the canonical entropy change for those initial phase points $\xib^0$ which happen
to lead to
the same work as the average work in the fictitious canonical ensemble, $W(\xib^0)=\W$.

\subsection{A theorem} 
As our first main result,
we  prove 
\beq
\DS\geq 0.
\ee 
Consider two normalized distributions $p_0(\xib^0)$ and $p_1(\xib^0)$ that
both vanish nowhere on phase space. Set $x=p_1/p_0$ in the trivial $\ln (1/x)\geq 1-x$ and average
both sides using $p_0$. This leads to 
\beq
\int d\xib^0 p_0(\xib^0) \ln[p_0(\xib^0)/p_1(\xib^0)] \geq 0
\label{eq:dkl} 
\ee known as positivity of the relative entropy.
We choose 
\beq
p_0(\xib^0)=\exp[-\beta_0 (H(\xib^0,\lb^0)-F(\beta_0,\lb^0))]
\ee
and
\beq 
p_1(\xib^0)=\exp[-\beta_1(H(\xib^1(\xib^0),\lb^1) -F(\beta_1,\lb^1))]
\ee 
with yet arbitrary $\beta_0$ and $\beta_1$.
Both distributions are normalized where for $p_1$ we exploit Liouville's theorem.
Inserting them into the inequality (\ref{eq:dkl})  leads to
\beq
\beta_1[E^0+\W_0-F(\beta_1,\lb^1)]\geq \beta_0[E^0-F(\beta_0,\lb^0)]
\label{eq:theo}
\ee
valid for any $\beta_0$ and $\beta_1$ with $\W_0$ the mean work associated with the
driving $\lb^\tau$ for an initial distribution drawn from a fictitious canonical
ensemble at $(\beta_0,\lb^0)$. Specializing to
\beq
\beta_0 =\bb(E^0,\lb^0), 
\ee which implies $\W_0=\W$, and to
\beq
\beta_1 = \bb(E^0+\W,\lb^1)
\ee
proves $\DS\geq 0$.
 
If one could replace the fluctuating work in the micro-canonical distribution by
the corresponding mean canonical one, that is by $\W$, a second law for the entropy change
were proven through the theorem. While such a replacement might seem plausible
for $N\to\infty$, a more detailed analysis is warranted.

\subsection{Fluctuations in the large-$N$ limit}

We first consider a system that is non-infinitesimally driven, i.e., the control
parameter changes in finite time by an order one and the driving affects a non-negligible part
of the system. The sign and magnitude of the terms in the series (\ref{eq:exp}) can then be estimated as follows.
(i) 
$\DS$ is non-negative because of the theorem. For a macroscopic system 
driven as just described, $\DS$ will typically be of the order of the 
number of degrees 
of freedom $N$. (ii) For such a process, the average of the
micro-canonical work, $\langle W(\xib^0)\rangle $, and the mean canonical work, $\W$, will typically be both of order $N$.
Their difference $\langle \delta W\rangle$ should then be of order one. 
This is most obvious for a quench, $\lb^0\to\lb^1$,
since the work is then given by the corresponding mean values of the phase
space function $H(\bxi,\lb^1)-E^0$. For such an extensive observable, canonical and
micro-canonical average differ by order one
as shown in the Appendix. (iii) In the sum, for $k=2$,
$\langle (\delta W)^2 \rangle$ will typically scale $\sim N$. Since, $\partial_E\bb\sim (-)1/N$,
in the sum, the leading term with $k=2$ will be negative of order 1. 

Under these assumptions, the sum of second and third term in (\ref{eq:exp}) is at most negative
of order 1. The first term, $\DS$, is definitely non-negative and typically positive of order $N$.
Contributions involving the higher order moments $\langle \delta W(\xib^0)^k\rangle, k\geq 3$ vanish
for large $N$.  For a macroscopic system, we have thus shown that, on the scale $N$, the canonical entropy for a
driven, 
thermally isolated system does not decrease on average.
However, the
 possible presence of non-extensive negative terms of $O(1)$
prevents us from proving a sharper, strict second law
in the  form $\langle \Delta \bS\rangle \geq 0$ for a
finite system.

For an infinitesimal quench of order $\delta \lb$ affecting a large system, the scaling of the first three
terms in the expansion (\ref{eq:exp}) with $\delta \lambda$ can be inferred as follows while their scaling with $N$ as just derived remains unchanged. The first term is non negative and
hence will scale $\sim N(\delta \lambda)^2$. The second one can be of
any sign and hence is linear in $\delta \lambda$. In the third one, since $\delta W\sim \delta\lambda$, we get $\langle (\delta W)^2\rangle \sim (\delta\lambda)^2$. Summarizing, for small $\delta\lambda$, we have 
\beq
\langle \Delta S_C\rangle \approx c_1 N (\delta \lb)^2\pm c_2\delta\lb - c_3 (\delta \lb)^2,
\ee
with positive constants $c_{1,2,3}$ of order 1. Consequently, there will generically
be a one-sided small range $|\delta \lb|\lsim 1/N$ for which the mean change in
canonical entropy will be negative of order $-1/N$ for a large system.

\section{Standard micro-canonical entropies}
We now turn to the status of the second law for a driven system  using the
two more established variants
of the micro-canonical entropy, the Boltzmann entropy and the Gibbs volume entropy.
 The Boltzmann entropy is defined as 
\beq
\Smic(E,\lb)\equiv \ln [\delta E\int d\xib~\delta(E-H(\xib,\lb))] ,
\ee
where $\delta E$ is a constant parameter to render the logarithm dimensionless. It will drop out in the
final expressions. The volume, or Gibbs entropy, is defined by 
\beq
\Sg(E,\lb)\equiv \ln [\int d\xib~\theta(E-H(\xib,\lb))] .
\ee 
It will be convenient to define the Gibbs temperature
\beq
\beta_G(E,\lb)\equiv \partial_E \Sg(E,\lb)
\ee
implying 
\beq
\Smic=\Sg+\ln(\delta E~\beta_G) .
\label{eq:bg} 
\ee

We first recall the derivation of the micro-canonical Crooks relation \cite{cleu06a}
on which we will base the further analysis. The equations of motion are reversible which
implies that
\beq
W(\xib^0)=-\tilde W(\tilde \xib^0)
\ee
where $\tilde W(\tilde \xib^0)$ is the work spent in the reverse process driven by the 
reversed protocol
$\tilde\lb^\tau\equiv \lb^{t-\tau}$ and starting at $\tilde \xib^0$, defined as
 the final point 
$\xib^1(\xib^0)$ of the original process with reversed momenta. Consequently, 
\beq
\delta(W(\xib^0)-W)=\delta(\tilde W(\tilde \xib^0)+W) .
\ee
Integrating this relation over the energy shell at $E^0$, yields the micro-canonical Crooks relation \cite{cleu06a}
\begin{eqnarray}
p(W|E^0)e^{\Smic(E^0,\lb^0)}
&=&  \int d\xib^0\delta(H(\xib^0,\lb^0)-E^0)\delta(\tilde W(\tilde \xib^0)+W)\nn\\
&=&\int d\xib^1\delta(H(\xi^1,\lb^1)-E^0-W)\delta(\tilde W(\tilde \xib^0)+W)\nn\\
&=&\tilde p(-W|E^0+W)e^{\Smic(E^0+W,\lb^1)} .
\label{eq:mc}
\end{eqnarray}
Here, $p(W|E)$ and $\tilde p(W|E)$ are the micro-canonical distributions on the energy shell
$E$ for a work $W$ for the original and the reversed process, respectively.
For the change in Boltzmann entropy
\beq
\Delta\Smic(\xib^0)\equiv \Smic(E^0+W(\xib^0),\lb^1) - \Smic(E^0,\lb^0) ,
\label{eq:Smix}
\ee one finds, 
 after getting
$e^{\Smic(E^0+W,\lb^1)}$ to the left hand side and integration over $W$,
the exact result
\beq
\langle e^{-\Delta \Smic} \rangle =\int dW \tilde p(-W|E^0+W) .
\ee Because of the conditioning on $E^0+W$ the integral is not 1, in general. Consequently,
one cannot infer $\langle \Delta \Smic\rangle \geq 0$, in general, either. To proceed, we apply an expansion on the right hand
side, leading to
\beq
\langle e^{-\Delta \Smic} \rangle =1+\sum_{k=1}^\infty\frac {(-1)^k}{k!}\partial^k_E\langle \tilde W^k\rangle,
\label{eq:wmic}
\ee
where $\tilde W$ is the work in the reversed process and the average is over the energy shell $E^0$.
This original  result is still formally exact.

We now consider a large system, for which the driving affects a substantial
fraction of all degrees of freedom. We then expect $\tilde W$ to scale like $E$,
\beq
\langle \tilde W\rangle \approx \tilde\alpha E,
\label{eq:alpha}
\ee with $\tilde \alpha$ of order 1 being a functional of the driving protocol.
Likewise, we expect the  variance of the work to be of order $E$ and a corresponding scaling of the higher cumulants. Under these assumptions the terms in the series (\ref{eq:wmic}) become
\beq
\partial^k_E \langle \tilde W^k\rangle\approx \partial^k_E \langle \tilde W\rangle^k\approx k!\tilde\alpha^k .
\ee Inserted in (\ref{eq:wmic}) and summing the geometrical series leads to
the integral fluctuation relation 
\beq
\langle e^{-\Delta \Smic}\rangle = 1/(1+\tilde\alpha)
\label{eq:ift}
\ee for a large driven system. Jensen's inequality then implies
\beq
\langle \Delta \Smic \rangle \geq \ln(1+\tilde\alpha).
\label{eq:ineq}
\ee
Since $\tilde\alpha$ can be negative, as exemplified below, the average 
change in Boltzmann
entropy is, even for a large system, not necessarily positive. However, as above 
for the change in
canonical entropy, it cannot
be negative of order $E$, i.e., of order $N$.

For the change in Gibbs entropy
\beq
\Delta\Sg(\xib^0)\equiv \Sg(E^0+W(\xib^0)),\lb^1) - \Sg(E^0,\lb^0) ,
\label{eq:Sgm}
\ee
 a similar result can be derived as follows.
With (\ref{eq:bg}), and (\ref{eq:mc}), we
obtain \begin{eqnarray}
\langle e^{-\Delta \Sg}\rangle &=& \int dW p(W|E^0) e^{-\Delta \Sg} \nn\\
&=& \int dW \tilde p(-W|E^0+W)e^{(\Delta \Smic-\Delta \Sg)}\nn\\
&=&\frac{1}{\beta_G(E^0,\lb^0)} \int dW \tilde p(-W|E^0+W) \beta_G(E^0+W),\lb^1)\nn\\
&=&\frac{1}{\beta_G(E^0,\lb^0)}\left[\beta_G(E^0,\lb^1)+
\sum_{k=1}^\infty\frac {(-1)^k}{k!}\partial^k_E[\langle \tilde W^k\rangle \beta_G(E,\lb^1)]_{|E=E^0}\right] .
\end{eqnarray} Since $\beta_G(E,\lb)\approx c N/E$,
 where $c$ is of
order 1, each term in the sum will vanish in the limit of a large system if the scaling (\ref{eq:alpha}) is assumed as
above. Consequently, in this limit,
\beq
 \langle e^{-\Delta \Sg}\rangle =\beta_G(E^0,\lb^1)/\beta_G(E^0,\lb^0) .
\ee
The corresponding inequality becomes 
\beq
\langle \Delta \Sg\rangle \geq \ln [\beta_G(E^0,\lb^0)/\beta_G(E^0,\lb^1)].
\label{eq:Sg}
\ee For a large system,
the right hand side will be of order 1 but not necessarily positive. As in the
two cases discussed above, we find that for a large system, the mean change in Gibbs volume
entropy cannot be extensively negative, i.e., negative of order $N$.

\section{Illustrative examples}
These results can be illustrated using exactly solvable models.

\subsection{Quenching an $N$-dimensional  harmonic oscillator}
 We 
consider an isotropic $N$-dimensional harmonic oscillator ($\xib\equiv\{q_j,p_j\}, j=1,...,N$) 
with mass $m$ and energy $E^0$ that is 
instantaneously quenched from 
an initial stiffness $\l^0$ to a final one 
\beq
\l^1\equiv \gamma \l^0.
\ee
 The work spent in
this process can be written as
\beq
W(\xib^0)=(\gamma-1)x(\xib^0)E^0
\ee
where 
\beq
x(\xib^0)\equiv \l^0\sum_{j=1}^N (q_j^0)^2/(2E^0)
\ee denotes that fraction of the total energy which the potential energy
carries initially ($0\leq x\leq 1$). The micro-canonical distribution $p(x)$ can be
calculated exactly and is given by the  symmetric beta distribution
\beq
p(x)=c_N [x(1-x)]^{N/2-1}
\label{eq:px}
\ee
with mean $
\langle x\rangle =1/2$ and  normalization $c_N\equiv \Gamma(N)/\Gamma^2(N/2)$ involving the
Gamma function.

The change in Boltzmann entropy
(\ref{eq:Smix}) becomes
\beq
\Delta \Smic(x) = \ln\frac{[1+(\gamma-1)x]^{N-1}}{\gamma^{N/2}}  .
\label{eq:dsmic}
\ee Its mean
$\langle \Delta S_B\rangle =\int_0^1dx p(x) \Delta S_B(x)$ is  negative for
$N=1,2$ and any $\gamma>1$. For
any $N\geq 3$, there exists a range $1<\gamma<\gamma^*(N)$ 
for which $\langle \Delta S_B \rangle <0$. For $N\to \infty$, $\gamma^*(N) \to 1$.

The canonical entropy of the $N$-dimensional oscillator is
given by
\beq
\bS(E,\l)=\bb[E-F(\bb,\l)]=N\left[1+\ln\frac{2\pi m^{1/2} E}{N h \l^{1/2}}\right]
\ee using $\bb(E,\l)=N/E$, and $h$ Planck's constant. Consequently, 
the change in canonical entropy after the quench from $\l^0$ to $\l^1$ 
becomes
\beq
\Delta \bS(x)=N\ln\frac{1+(\gamma-1)x}{\gamma^{1/2}} ,
\ee which is not necessarily positive. However,
the mean change $
\langle \Delta \bS\rangle=\int_0^1 dx p(x) \Delta \bS(x)
$ is positive  for any $\gamma\not=1$ and any $N$. 
We also get
\beq
\DS=\Delta S_C(x=1/2)=N\ln\frac{1+\gamma}{2\gamma^{1/2}} \geq \langle \Delta \bS\rangle \geq 0 .
\ee This result illustrates the above discussion on the sign of the third
 term in (\ref{eq:exp}) since for this case the
second term vanishes due to  $\W=\langle W\rangle$.

For this quench from $\l^0 \to \l^1=\gamma \l^0$, the
change in Gibbs volume entropy is exactly the same as in canonical entropy, 
$\Delta \Sg(x)=\Delta \bS(x)$. Hence, the mean
change in volume entropy $\langle \Delta \Sg \rangle$  is positive for all $N$ as well. Moreover, since the Gibbs temperature
is independent of the stiffness at fixed energy,
as is the canonical temperature, the right hand side
of (\ref{eq:Sg}) 
vanishes thus confirming the strict positivity of $\langle \Delta \Sg\rangle$ in this
case for a large system.

\subsection{A two-dimensional harmonic oscillator quenched into a finite disc}

As a specific example for a mean negative entropy change for both the Gibbs volume
entropy  and the canonical one, consider a two-dimensional harmonic oscillator with microcanonical
initial conditions at energy $E^0$. At time $t=0$, a circular wall is introduced with
radius 
\beq
R=(2E^0 /\l^0)^{1/2} .
\ee Since this is the maximal possible elongation of the oscillator
from the origin, introducing this wall does not cost work. At the same time, the stiffness $\l^0$
is set to zero, thereby extracting work and thus lowering the energy of the system. 
For $t>0$ the system thus corresponds to a two dimensional particle
confined to a circular disc with final energy 
\beq
E^1(x)=x(\xib^0)E^0.
\label{eq:e1}
\ee In two-dimensions, the
distribution $p(x)=1$ is constant as given by (\ref{eq:px})
for $N=2$.

Initially, the Gibbs volume entropy  is
\beq
S_G(E^0)=\ln\left[ \frac{(2\pi)^2}{h^2}\int_0^\infty dp p \int_0^\infty dq q \theta((E^0)^2-p^2/2m-\l^0q^2/2)\right]
=
\ln
\frac{2\pi^2(E^0)^2m}{\l^0h^2}. 
\ee
Depending on the final energy (\ref{eq:e1}), the final Gibbs volume entropy
becomes 
\beq
S_G(E^1(x))=\ln \left[\frac{(2\pi)^2}{h^2}\int_0^\infty dp p \int_0^{R}dq q \theta(E^1-p^2/2m)\right]
=
\ln
\frac{2\pi^2E^1(x)mR^2}{h^2}. 
\ee
Consequently, the mean entropy change becomes
\beq
\langle \Delta S_G(x)\rangle = \int_0^1 dx \ln (2x)=\ln 2 -1<0.
\ee

For the canonical entropy, we have initially with $\bb=2/E^0$ 
\beq
S_C(E^0)= 2 +\ln\frac{(2\pi)^2m(E^0)^2}{4\l^0 h^2}.
\ee After the quench, the particle in the two-dimensional disc has
an entropy
\beq
S_C(E^1) = 1+\ln\frac{2\pi^2mE^1 R^2}{h^2} ,
\ee where we use
 $\bb^1=1/E^1$.
The mean entropy change thus becomes
\beq
\langle \Delta S_C\rangle = -1 +\int_0^1dx \ln(4x)=-2+2\ln2<0 .
\ee
For both entropy versions, the mean entropy change is thus negative in this quench. 
Further examples with a mean negative entropy change can be constructed \cite{fogl18}.

\section{Concluding summary} 

We have investigated whether a second law for the mean entropy change of an
isolated driven system can be proven.
For three variants of  entropy (canonical, Boltzmann, and Gibbs volume) starting from micro-canonical initial
conditions  for a large system, we have shown that 
the average entropy change cannot be negative of order $N$. 
This result was based on a theorem for the canonical entropy and on 
explicit series expressions derived from the micro-canonical Crooks relation
for the two standard variants, using in all three cases rather mild assumptions on
the scaling of the moments of the work distribution. 
For a large but finite system, however, the mean entropy
change
can very well be negative as it has 
been demonstrated explicitly for all three versions of entropy. 

Concerning future perspectives, this study first prompts  the quest for a non-existence proof of 
an entropy function $S(E,\l)$ whose mean change is non-negative for arbitrary driving $\l^\tau$ in any finite system. Since the three, arguaby most prominent, candidates of such
a function have been excluded here through counter-examples, it is unlikely that any other
reasonable variant will
behave qualitatively differently. Second, it would be interesting to explore more generally
bounds on the maximal possible negative values that can be reached for the mean entropy
change in the large-$N$ limit. For infinitesimal changes of a control parameter,
our series expansion for canonical entropy suggest that negative values of order $-1/N$ are generic. The
arguments  based on the integral fluctuations relations for Boltzmann and 
Gibbs entropy given here cannot not even exclude the possibility of negative values of order 1
for finite quenches.

\appendix

\section{N-dependence of the difference between  canonical and microcanonical mean values}

\def\xh{\hat x}
In the micro-canonical ensemble of energy $E$, the mean value of an observable $A$ is given by 
\beq
\langle A|E\rangle = \int d\xi\delta(H(\xi)-E)A(\xi)\delta E 
e^{-S_B(E)} ,\ee
where the Boltzmann entropy
\beq
S_B(E)\equiv [\ln \delta E\int d\xi \delta(H(\xi)-E)] 
\ee
with an irrelevant constant parameter $\delta E$ of dimension energy provides the normalization.

The mean value of an observable $A(\xi)$ in the
canonical ensemble at inverse temperature $\beta$ and free energy
$F(\beta)$ can then be expressed as
\begin{eqnarray}
\langle A|\beta\rangle &=& \int d\xi\exp[-\beta(H(\xi)-F(\beta))]A(\xi)\nn\\
&=& \int dE \int d\xi\exp[-\beta(H(\xi)-F(\beta))]A(\xi)\delta(H(\xi)-E)\nn\\
&=& \frac{\int dE e^{-\beta E+S_B(E)}\langle A|E\rangle}
{\int dE e^{-\beta E + S_B(E)}},
\label{eq:int}
\end{eqnarray}

We now perform a saddle point evaluation of both integrals
in (\ref{eq:int}) around $\hat E$ implicitly defined through
\beq
\beta=\partial_ES_B(E)_{|E=\hat E}\equiv  \bbb(\hat E).
\label{eq:hatebb2}
\ee
In the nominator, we have to expand $\langle A|E\rangle$ to second
order  and the exponent to third order in $y \equiv E-\hat E$. The final result is\footnote{For a function
$f(x)$ with a narrow  maximum at  $\xh$, hence $f''(\xh)<0$, a function $g(x)$, and $y\equiv x-\xh$, we use
\begin{eqnarray}
\int dx g(x)e^{f(x)}&\approx& \int dy g(\xh)\left[1+\frac{yg'(\xh}{g(\xh)}+\frac{y^2g''(\xh)}{2g(\xh)}\right]e^{f(\xh)+y^2f''(\xh)+y^3f'''(\xh)/6}\nn\\&\approx& g(\xh)e^{f(\xh)}\int dy \left[1+\frac{yg'(\xh)}{g(\xh)}+\frac{y^2g''(\xh)}{2g(\xh)}\right]\left[1+y^3f'''(\xh)/6\right]e^{y^2f''(\xh)/2}\nn\\
&=&g(\xh)[2\pi/|f''(\xh)|]^{1/2}\left[1+ \frac{g''(\xh)}{2g(\xh)|f''(\xh)|}+\frac{g'(\xh)f'''(\xh)}{2g(\xh)|f''(\xh)|^2}\right] .\nn
\end{eqnarray}
}
\beq
\langle A|\beta\rangle \approx \langle A|\hat E\rangle
\left[1
+\frac{\partial^2_E\langle A|E\rangle}{2|\partial_E \bbb|\langle A| E\rangle}
+\frac{\partial^2_E  \bbb\partial_E\langle A|E\rangle}{2|\partial_E \bbb|^2|\langle A|E\rangle} \right]_{|E=\hat E} .
\label{eq:final}
\ee 
The crucial point now is that in a large system $\partial_E\sim 1/N$.  Hence, the square bracket in (\ref{eq:final}) is $[1+O(1/N)]$. Consequently, the  canonical and the micro-canonical average of an
observable $A$ differ by a relative amount of order $1/N$. 

~~\\~\textbf{Dedication}\\

This paper is dedicated to the memory of Christian Van den Broeck,
pioneer in stochastic thermodynamics. His thought-provoking papers 
on the subject and the many inspiring and, reflecting his fine character, always friendly discussions we had over the last fifteen years strongly influenced my research in the field and my understanding of it. 

~~\\~~ \textbf{References}
~\\~
\bibliographystyle{elsarticle-num}
\bibliography{/home/public/papers-softbio/bibtex/refs} 

\end{document}